\documentclass{article}
\usepackage{ijcai25}

\usepackage[switch]{lineno}

\usepackage{hyperref}
\usepackage{url}
\usepackage{times}
\usepackage{latexsym}
\usepackage[T1]{fontenc}
\usepackage[utf8]{inputenc}
\usepackage{microtype}
\usepackage{inconsolata}
\usepackage{diagbox}
\usepackage{booktabs}
\usepackage{adjustbox}
\usepackage{verbatim}
\usepackage{color}
\usepackage{makecell}
\usepackage{multirow}
\usepackage{graphicx}
\usepackage{amsmath, amssymb, bm}
\usepackage{algorithm}
\usepackage{algorithmic}
\floatname{algorithm}{Algorithm}
\usepackage{color}

\title{A Survey on LLM-powered Agents for Recommender Systems}

\author{
    Qiyao Peng\textsuperscript{\rm 1} \and
    Hongtao Liu\textsuperscript{\rm 2} \and
    Hua Huang\textsuperscript{\rm 1} \and Qing Yang\textsuperscript{\rm 2} \and \And Minglai Shao\textsuperscript{\rm 1} \\
    \affiliations
    \textsuperscript{\rm 1}School of New Media and Communication, Tianjin University, Tianjin, China\\
    \textsuperscript{\rm 2}Du Xiaoman Financial Technology, Beijing, China\\
    \emails
    \{qypeng, huanghua18, shaoml\}@tju.edu.cn, \{liuhongtao01,yangqing\}@duxiaoman.com
}

\begin{document}

\maketitle

\begin{abstract}

% Traditional recommender systems, despite their widespread adoption in the digital era, continue to face fundamental challenges in understanding complex user intentions,enabling natural interactions, and providing explainable recommendations. 
% The emergence of Large Language Model (LLM)-powered agents has shown promising potential in addressing these limitations, leading to a new wave of research in recommender systems.
% This survey comprehensively examines the integration and development of LLM-powered agents in recommender systems.
% We systematically categorize existing research into three dominant paradigms: (1) Recommender-oriented method, focusing on utilizing different agents to tackle traditional recommendation tasks;  (2) Interaction-oriented method, focusing on enabling natural language interaction and enhancing recommendation interpretability through conversational engagement; and (3) Simulation-oriented method, focusing on employing various of agents to simulate user behaviors and item characteristics in recommender systems. 
% Besides, we provide an in-depth analysis of core components in LLM-powered agents for recommender systems, including: profile, memory, planning, and action modules.
% Then, we systematically review existing datasets and evaluation methodologies. 
% This survey not only presents the current landscape of LLM-powered agent recommender systems but also identifies key challenges and promising future research directions in this rapidly evolving field.

Recommender systems are essential components of many online platforms, yet traditional approaches still struggle with understanding complex user preferences and providing explainable recommendations.
The emergence of Large Language Model (LLM)-powered agents offers a promising approach by enabling natural language interactions and interpretable reasoning, potentially transforming research in recommender systems.
This survey provides a systematic review of the emerging applications of LLM-powered agents in recommender systems.
We identify and analyze three key paradigms in current research: (1) Recommender-oriented approaches, which leverage intelligent agents to enhance the fundamental recommendation mechanisms; (2) Interaction-oriented approaches, which facilitate dynamic user engagement through natural dialogue and interpretable suggestions; and (3) Simulation-oriented approaches, which employ multi-agent frameworks to model complex user-item interactions and system dynamics.
Beyond paradigm categorization, we analyze the architectural foundations of LLM-powered recommendation agents, examining their essential components: profile construction, memory management, strategic planning, and action execution. 
Our investigation extends to a comprehensive analysis of benchmark datasets and evaluation frameworks in this domain. 
This systematic examination not only illuminates the current state of LLM-powered agent recommender systems but also charts critical challenges and promising research directions in this transformative field.

\end{abstract}
\section{Introduction}

In the era of information explosion, recommender systems~\cite{wu2022survey} have become an indispensable component of digital platforms, helping users navigate through massive amounts of content across e-commerce, social media, and entertainment domains. 
While traditional recommendation approaches~\cite{he2017neural} have achieved considerable success in providing personalized suggestions through analyzing user preferences and historical behaviors, they still face significant challenges in real-world applications, such as limited understanding of complex user intents, insufficient interaction capabilities, and the inability to provide interpretable recommendations~\cite{zhu2024recommender}.

Recent advancements in Large Language Models (LLMs)~\cite{achiam2023gpt} have sparked increasing interest in leveraging LLM-powered agents~\cite{wang2024survey} to address the aforementioned challenges in recommender systems. 
The integration of LLM-powered agents into recommender systems offers several compelling advantages over traditional approaches~\cite{zhu2024recommender}. 
First, LLM agents can understand complex user preferences and generate contextual recommendations through their sophisticated reasoning capabilities, enabling more nuanced decision-making beyond simple feature-based matching. 
Second, their natural language interaction abilities facilitate multi-turn conversations that proactively explore user interests and provide interpretable explanations, enhancing both recommendation accuracy and user experience. 
Third, these agents revolutionize user behavior simulation by generating more realistic user profiles that incorporate emotional states and temporal dynamics, enabling more effective system evaluation. Furthermore, the pre-trained knowledge and strong generalization capabilities of LLMs facilitate better knowledge transfer across domains, addressing persistent challenges such as cold-start~\cite{shu2024rah} with minimal additional training.

In this survey, we present a comprehensive review of LLM-powered agents for recommender systems. 
First, we introduce the background of traditional recommender systems and discuss their limitations in understanding complex user intents, interaction capabilities, and interpretability. 
We then systematically examine how LLM-powered agents address these challenges through three main paradigms: recommender-oriented (e.g., \cite{wang2024recmind,wang2024macrec}), interaction-oriented (e.g., \cite{zeng2024automated,friedman2023leveraging}), and simulation-oriented (e.g., ~\cite{yoon2024evaluating,guo2024knowledge}) approaches. 
Following that, we propose a unified agent architecture consisting of four core modules (Profile~\cite{cai2024flow,zhang2024agentcf}, Memory~\cite{shi2024large,fang2024multi}, Planning~\cite{wang2023drdt,shi2024large}, and Action~\cite{cshi,toolrec}) and analyze how existing methods implement these components. 
Furthermore, we compile comprehensive comparisons of datasets (including Amazon series, MovieLens, Steam, etc.) and evaluation methodologies, encompassing both standard recommendation metrics and novel evaluation approaches. 
Finally, we explore several promising future directions in this field.
%To enhance the recommendation performance with LLM-powered agents, in this survey, we thoroughly examine existing methods and highlight three dominant paradigms in the literature. Methods falling under strategic recommendation [BLLP, DRDT] focus on optimizing decision-making and long-term planning through multi-agent collaboration and sequential reasoning. Interactive recommendation approaches [AutoConcierge, RecLLM] leverage the natural language capabilities of LLMs to enable dialogue-based interactions and provide explanatory feedback, thereby enhancing user engagement and system transparency. Examining user simulation methods [Agent4Rec, AgentCF] aids in understanding user behaviors and preferences by constructing artificial users with profile, memory, and action modules, which facilitates system evaluation and optimization. Some hybrid frameworks [FLOW, MACRec] combine multiple paradigms to achieve comprehensive improvements. For instance, FLOW integrates strategic planning with user simulation through a feedback loop mechanism, while MACRec combines multi-agent decision-making with interactive recommendation capabilities.

\begin{figure*}
    \centering
    \includegraphics[width=0.97\linewidth]{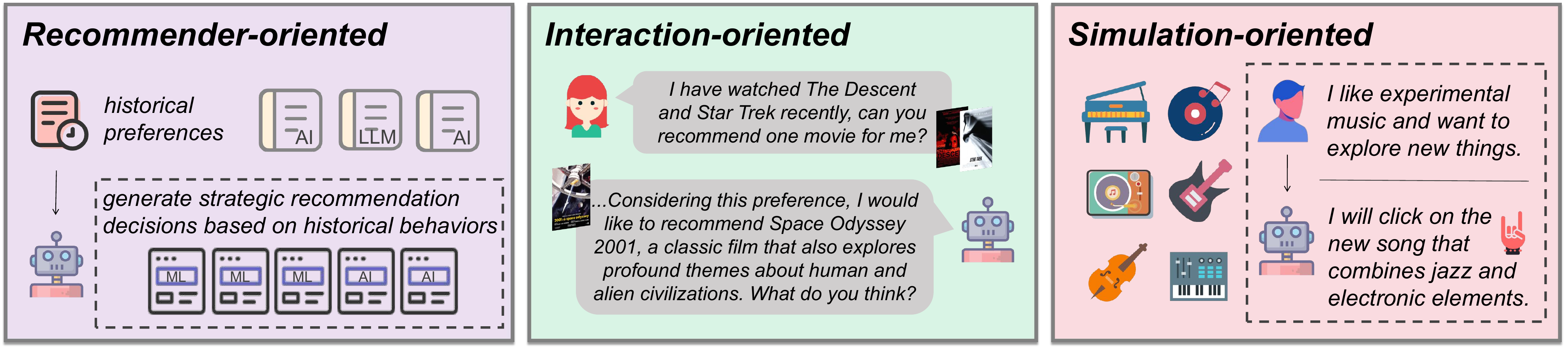}
    \caption{Illustration of Different Method Objectives. We classify existing methods into the following three categories: (1) Recommender-oriented method; (2) Interaction-oriented method; (3) Simulation-oriented method.}
    \label{objective}
\end{figure*}

\begin{itemize}
    %\item We systematically categorize LLM-powered agent for recommender systems into three paradigms (recommender-oriented, interaction-oriented, and simulation-oriented).
    \item We propose  a systematic categorization of LLM-powered recommender agents, identifying three fundamental paradigms: recommender-oriented, interaction-oriented, and simulation-oriented approaches. This taxonomy provides a structured framework for understanding current research.

    %\item We analyze existing recommender systems through the lens of a unified agent architecture framework, which consists of four core modules: Profile, Memory, Planning, and Action. This analysis explores how different methods combine these components.
    \item We utilize a unified architectural framework for analyzing LLM-powered agent recommender, decomposing them into four essential modules: Profile Construction, Memory Management, Strategic Planning, and Action Execution. Through this framework, we systematically examine how existing methods integrate and implement these components.

    %\item We compile comprehensive comparisons of existing methods, datasets (including Amazon series, MovieLens, Steam, etc.), and evaluation methodologies (from standard recommendation metrics to novel evaluation approaches).
    \item We provide a comprehensive comparative analysis of existing methods, benchmark datasets, and evaluation methodologies, encompassing both traditional recommendation metrics and emerging evaluation approaches specifically designed for LLM-powered agent recommender.

    %\item We explore several promising  directions in LLM-powered agent for recommender systems for the future works.
\end{itemize}

\begin{figure}
    \centering
    \includegraphics[width=0.95\linewidth]{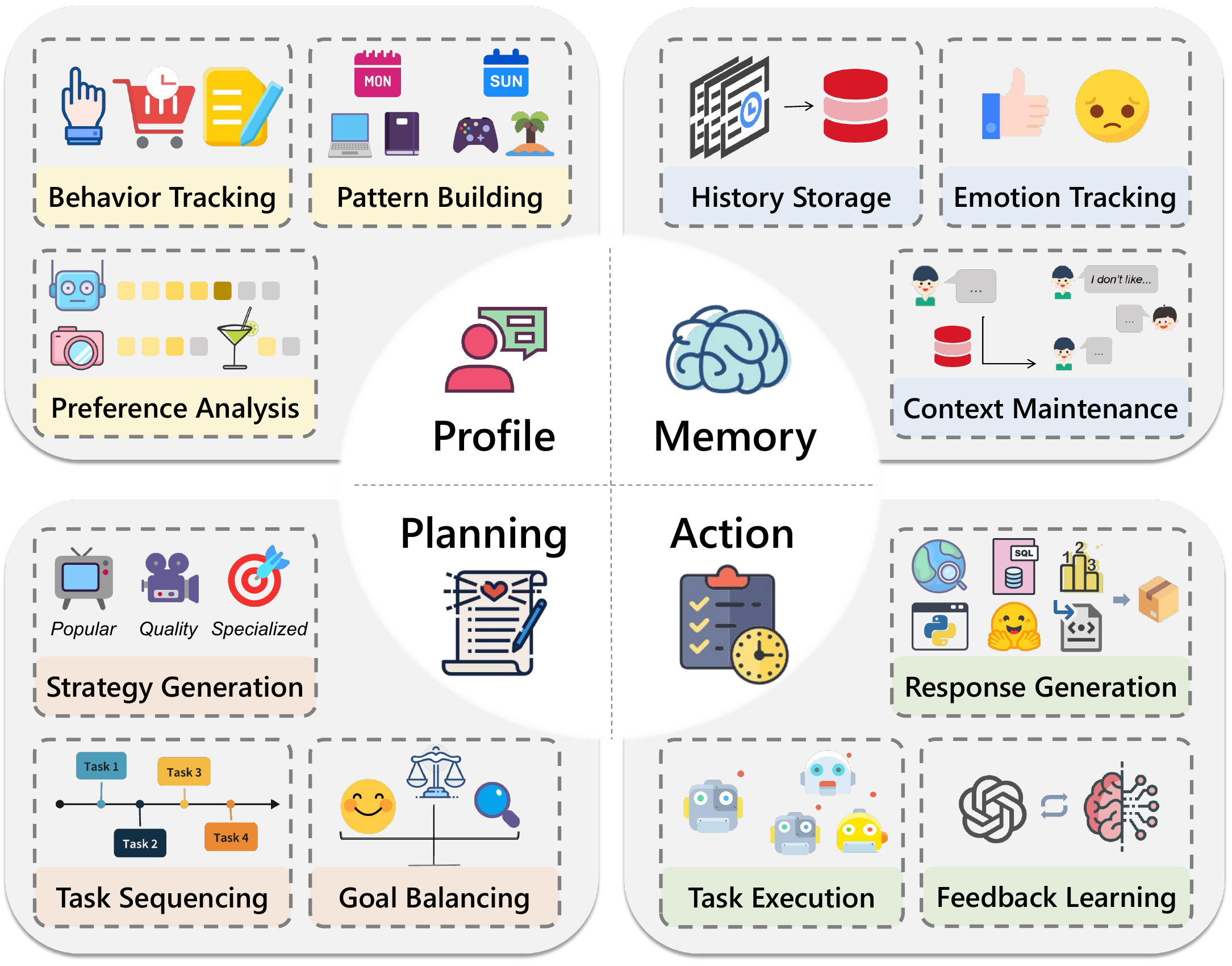}
    \caption{Illustration of Agent Components and Corresponding Functions.}
    \label{component}
\end{figure}
\section{Background}

\subsection{Traditional Recommendation} 

In conventional recommendation systems, the problem is typically formulated over a user space $\mathcal{U} = {[u_1, u_2, ..., u_m]}$, an item space $\mathcal{I} = {[i_1, i_2, ..., i_n]}$, and their interaction matrix $\mathcal{D} \in \mathbb{R}^{m \times n}$. The fundamental goal is to learn a preference function $p: \mathcal{U} \times \mathcal{I} \rightarrow \mathbb{R}$ that predicts user preferences:
\begin{equation}
\min_{\theta} \sum_{(u,i) \in \mathcal{D}} \mathcal{L}(p_{\theta}(u,i), y_{u,i}) \ ,
\end{equation}
where $p_{\theta}(u,i)$ represents the predicted preference and $y_{u,i}$ denotes the ground truth interaction. 
While various approaches have been proposed, from matrix factorization~\cite{hu2008collaborative} to deep learning~\cite{he2017neural}, these traditional methods face several inherent limitations. 
First, they struggle to understand complex user intents beyond numerical interactions. 
Second, they lack the ability to engage in meaningful interactions to explore user preferences. 
Third, their recommendations often appear as a ``black box'' without clear explanations for users.

\subsection{LLM as Agent}

Large Language Model (LLM) as an agent is an emerging research direction that has garnered significant attention~\cite{park2023generative}. 
By transcending the traditional static prompt-response paradigm, it establishes a dynamic decision-making framework~\cite{patil2023gorilla} capable of systematically decomposing complex tasks into manageable components. 
A typical LLM-powered agent architecture integrates four fundamental modules~\cite{wang2024survey}: (1) the Profile module, which constructs and maintains comprehensive user feature representations; (2) the Memory module, which orchestrates historical interactions and preserves contextual information for systematic experience accumulation; (3) the Planning module, which formulates strategic policies through sophisticated task decomposition and multi-objective optimization; and (4) the Action module, which executes decisions and facilitates environment interaction. 
The emergence of pioneering works such as ReAct~\cite{react}, Toolformer~\cite{toolformer}, and HuggingGPT~\cite{hugginggpt} has significantly advanced this field.

\subsection{LLM Agents for Recommendation}

In LLM-powered agent for recommender systems, we formulate the recommendation process through an agent-centric framework. 
Let $a \in \mathcal{A}$ denote an agent equipped with a set of functional modules $\mathcal{F} = {\mathcal{F}_1, \mathcal{F}_2, ..., \mathcal{F}_K}$, where each module $\mathcal{F}_k$ represents a specific capability. 
The recommendation process for a user $u$ can be formally expressed as:
\begin{equation}
\hat{\mathbf{y}}_u = f({\mathcal{F}_k(X_u)}), k=1 \cdots K \ ,
\end{equation}
where $X_u \in \mathcal{X}$ represents the input space containing user-specific information (e.g., interaction history, contextual features), and $\hat{\mathbf{y}}_u \in \mathbb{R}^N$ denotes the predicted preference distribution over the item space. 
The integration function $f: {\mathcal{F}_k(X_u)} \rightarrow \mathbb{R}^N$ synthesizes module outputs to generate final recommendations.
Building upon the previously introduced four functional module (Profile, Memory, Planning, and Action), this formulation provides a flexible framework that can accommodate various LLM-powered agent recommendation approaches. 
These modules operate in a closed-loop framework, where interaction data continuously enriches user profiles and system memory, informing planning strategies that ultimately manifest as personalized recommendations through action execution and feedback collection.
\section{Methods}

In this section, we sort out existing LLM-powered agent recommendation works based on the overall objective of the method and the agent components of different methods.

\subsection{Method Objective}

In Table~\ref{tab:comparison}, we classify method objectives of existing methods into three categories: recommender-oriented approaches, interaction-oriented methods, and simulation-oriented methods.
The illustrations of categories are shown in Figure~\ref{objective}.

\textbf{(1) Recommender-oriented} approaches focus on developing intelligent recommendation equipped with enhanced planning, reasoning, memory, and tool-using capabilities. 
In these approaches, LLMs leverage users' historical behaviors to generate direct recommendation decisions. 
For instance, as shown in Figure~\ref{objective}, when a user demonstrates recent engagement with technology news and AI-related content, the system might strategically recommend: ``Here are 5 articles about latest large language model breakthroughs, 3 introductory articles about machine learning basics, and 2 popular science pieces about AI's impact on society.''
This paradigm demonstrates how agents can effectively combine their core capabilities to deliver direct item recommendations.

Representative works in this direction include RecMind~\cite{wang2024recmind}, which develops a unified LLM agent with comprehensive capabilities to generate recommendations directly through LLM outputs.
MACRec, which introduces an agent-collaboration mechanism that orchestrates different types of agents to provide personalized recommendations~\cite{wang2024macrec}.

\textbf{(2) Interaction-oriented} methods focus on enabling natural language interaction and enhancing recommendation interpretability through conversational engagement. 
These approaches utilize LLMs to conduct human-like dialogues or explanation while making recommendations. 
For example, as shown in Figure~\ref{objective}, an LLM might respond to a user query with: ``I noticed that you like science fiction movies, especially after watching The Descent and Star Trek recently. Considering this preference, I would like to recommend Space Odyssey 2001, a classic film that also explores profound themes about human and alien civilizations. What do you think?''
Such interactive recommendations showcase the agent's ability to not only track user preferences but also articulate recommendations in a conversational manner, explaining the reasoning behind suggestions.

AutoConcierge~\cite{zeng2024automated} uses natural language conversations to understand user needs and collect user preferences, and uses LLM to understand and generate language, ultimately providing explainable personalized restaurant recommendations.
RAH~\cite{shu2024rah} is a human-computer interaction recommendation framework based on LLM agents. 
It realizes personalized recommendations and user intent understanding through the ResSys-Assistant-Human tripartite interaction and the Learn-Act-Critic loop mechanism.

\begin{table*}[t]
\resizebox{0.92\linewidth}{!}{
\begin{tabular}{c|c|c|c|c|c}
\toprule[1.5pt]
\textbf{Category} & \textbf{Methods} & \textbf{Profile Module} & \textbf{Memory Module} & \textbf{Planning Module} & \textbf{Action Module} \\
\midrule
\multirow{7}{*}[-1.5em]{\textbf{\makecell*[c]{Recommender- \\ oriented \\ Method}}} 
 & RAH~\cite{shu2024rah}  & $\times$ & \checkmark & \checkmark & \checkmark \\
\cmidrule{2-6}
 & ToolRec~\cite{toolrec} & $\times$  & \checkmark & $\times$ & \checkmark \\
\cmidrule{2-6}
 & PMS~\cite{pms} & \checkmark  & $\times$ & $\times$ & \checkmark  \\
\cmidrule{2-6}
 & DRDT~\cite{wang2023drdt} & $\times$ & $\times$ & \checkmark & $\times$  \\
\cmidrule{2-6}
 & BiLLP~\cite{shi2024large} & $\times$ & \checkmark & \checkmark & \checkmark  \\
\cmidrule{2-6}
 & RecMind~\cite{wang2024recmind} & $\times$ & \checkmark & \checkmark & \checkmark  \\
\cmidrule{2-6}
 & MACRec~\cite{wang2024macrec} & \checkmark & $\times$ & \checkmark & \checkmark  \\
\midrule
\multirow{7}{*}[-1.5em]{\textbf{\makecell*[c]{Interaction- \\ oriented \\ Method}}} 
& AutoConcierge~\cite{zeng2024automated} & $\times$ & \checkmark & \checkmark & \checkmark  \\
\cmidrule{2-6}
& MACRS~\cite{fang2024multi} & \checkmark & \checkmark & \checkmark & \checkmark \\
\cmidrule{2-6}
 & RecLLM~\cite{friedman2023leveraging} & \checkmark & \checkmark & $\times$ & \checkmark  \\
\cmidrule{2-6}
 & InteRecAgent~\cite{huang2023recommender} & \checkmark & \checkmark & \checkmark & \checkmark  \\
\cmidrule{2-6}
 & MAS~\cite{thakkar2024personalized} & \checkmark & \checkmark & \checkmark & \checkmark \\
\cmidrule{2-6}
 & H-MACRS~\cite{hmacrs} & \checkmark  & \checkmark & $\times$ & \checkmark  \\
\cmidrule{2-6}
 & Rec4Agentverse~\cite{Rec4Agentverse} & \checkmark  & $\times$ & \checkmark & $\times$  \\
\midrule
\multirow{8}{*}[-2em]{\textbf{\makecell*[c]{Simulation- \\ oriented \\ Method}}} & KGLA~\cite{guo2024knowledge} & \checkmark & \checkmark & $\times$ & \checkmark \\
\cmidrule{2-6}
 & CSHI~\cite{cshi} & \checkmark & \checkmark & $\times$ & \checkmark \\
\cmidrule{2-6}
 & SUBER~\cite{suber} & \checkmark  & \checkmark & $\times$ & $\times$ \\
\cmidrule{2-6}
 & LUSIM~\cite{lusim} & \checkmark  & \checkmark & $\times$ & $\times$ \\
\cmidrule{2-6}
 & FLOW~\cite{cai2024flow} & \checkmark & \checkmark & $\times$ & \checkmark \\
\cmidrule{2-6}
 & Agent4Rec~\cite{zhang2024generative} & \checkmark & \checkmark & $\times$ & \checkmark \\
\cmidrule{2-6}
 & AgentCF~\cite{zhang2024agentcf} & \checkmark & \checkmark & $\times$ & \checkmark  \\
\cmidrule{2-6}
 & UserSimulator~\cite{yoon2024evaluating} & \checkmark & $\times$ & $\times$ & \checkmark  \\
 \cmidrule{2-6}
 & RecAgent~\cite{wang2023user} & \checkmark & \checkmark & $\times$ & \checkmark  \\
\bottomrule[1.5pt]
\end{tabular}}
\centering
\caption{Comparison of Different LLM-powered Agent Recommendation Methods.}
\label{tab:comparison}
\end{table*}

\textbf{(3) Simulation-oriented} methods aim to authentically replicate user behaviors and preferences through sophisticated simulation techniques. These approaches leverage LLMs to generate realistic user responses to recommendations. For instance, when simulating user feedback, an LLM might generate: ``As a user who is keen to explore new music, I will click on this new song that combines jazz and electronic elements because it matches my interest in experimental music while maintaining the rhythmic style that I like.''
These methods focus on using agents to simulate user behaviors and item characteristics in RSs.

Agent4Rec~\cite{zhang2024generative} utilizes LLM-empowered generative agents as user simulators to model authentic interactions between users and recommender systems, aiming to replicate and evaluate realistic user behaviors in recommendation environments.
AgentCF~\cite{zhang2024agentcf} models both users and items as LLM-powered agents that autonomously interact and collaboratively learn from each other to simulate authentic user-item interactions in recommender systems.
UserSimulator proposes~\cite{yoon2024evaluating} an evaluation protocol to assess LLMs as generative user simulators in conversational recommendation through five tasks to measure how closely these simulators can emulate authentic user behaviors.

\subsection{Agent Components}

The LLM-based agent recommendation architecture consists of four main modules: Profile Module, Memory Module, Planning Module, and Action Module.
Figure~\ref{component} illustrates the core components of the architecture and corresponding functions.

\textbf{(1) Profile Module} is a fundamental component that constructs and maintains dynamic representations of users and items in recommender systems. 
Through continuous analysis of historical interactions, it captures temporal and contextual patterns in user behavior. 
For example, when the system observes that a user often browses technology news on weekday mornings and likes to watch travel content on weekends, the Profile Module will build a user profile of ``focusing on technology news on weekdays and preferring leisure content on weekends''. 
This adaptive profiling approach integrates behavioral patterns, user preferences, and external knowledge to enable highly personalized recommendations.

The profile module in Agent4Rec~\cite{zhang2024generative} incorporates dual components: quantifiable social traits (activity, conformity, and diversity) and personalized preferences extracted via LLM, enabling a comprehensive simulation of user characteristics.
MACRec~\cite{wang2024macrec} incorporates a user and item analyst, which play a crucial role in understanding user preferences and item characteristics.
AgentCF~\cite{zhang2024agentcf} constructs natural language-based user profiles to capture dynamic user preferences and item profiles to represent item characteristics and potential adopters' preferences, enabling personalized agent-based collaborative filtering.

\textbf{(2) Memory Module} serves as a contextual brain that manages and leverages historical interactions and experiences to enhance recommendation quality. 
It maintains a structured repository of past interactions, emotional responses, and conversational context to enable more informed decisions. 
For example, in a restaurant recommendation scenario, when a user comments ``that Sichuan restaurant was too spicy last time'', the Memory Module retrieves the specific restaurant reference from historical interactions and incorporates this preference signal into future recommendations, helping avoid overly spicy options. 
Through this continuous accumulation and utilization of experiential knowledge, the module enables more personalized and context-aware recommendations that reflect users' past experiences and preferences.

RecAgent~\cite{wang2023user} comprises three hierarchical levels: sensory memory, short-term memory, and long-term memory. The sensory memory processes environmental inputs, while short-term memory serves as an intermediate layer that can be transformed into long-term memory through repetitive reinforcement. Long-term memory stores crucial reusable information and facilitates self-reflection and knowledge generalization.
Agent4Rec~\cite{zhang2024generative} consists of factual memory (recording interactive behaviors) and emotional memory (capturing psychological states), stored in both natural language and vector representations, and managed through three mechanisms: retrieval, writing, and reflection.

\textbf{(3) Planning Module} outputs intelligent recommendation strategies by designing multi-step action plans that balance immediate user satisfaction with long-term engagement goals. 
It dynamically formulates recommendation trajectories through careful strategy generation and task sequencing. 
For example, in video recommendation, the system might construct a strategic plan: ``first recommend a popular video to establish user interest, and then gradually introduce niche but high-quality related content, while maintaining the diversity of genres, and ultimately achieve the goal of both satisfying user interest and expanding horizons''.
Through this planning approach, the module optimizes resource allocation and adapts recommendation sequences to achieve both user engagement and item discovery.

BiLLP~\cite{shi2024large} planning mechanism employs a hierarchical structure with two levels: macro-learning (Planner and Reflector LLMs) generates high-level strategic plans and guidelines from experience, while micro-learning (Actor-Critic) translates these plans into specific recommendations.
MACRS~\cite{fang2024multi} uses a multi-agent planning system where a Planner Agent coordinates three Responder Agents (Ask, Recommend, Chat) through multi-step reasoning. 
The system adjusts its dialogue strategy through a feedback mechanism, enabling reflective planning based on user interactions.

\textbf{(4) Action Module} serves as the execution engine that transforms decisions into concrete recommendations through systematic interaction with various system components.  
For example, in an e-commerce scenario, when receiving the directive ``recommend entry-level camera for new user'' from the Planning Module, the Action Module executes a coordinated sequence: analyzing purchase patterns of similar users, querying the product database with specific price and feature constraints, generating targeted recommendations, and capturing user feedback. 
This execution enables the system to deliver contextually appropriate recommendations while continuously learning from interaction outcomes.

RecAgent~\cite{wang2023user} orchestrates naturalistic agent interactions within recommender systems and social environments through a unified prompting framework, incorporating six action modalities (encompassing search, browse, click, pagination, chat, and broadcast functionalities).
InteRecAgent~\cite{huang2023recommender} action module integrates three core tools (information querying, item retrieval, and item ranking) while leveraging a Candidate Bus for sequential tool communication, enabling an end-to-end interactive process from user queries to final recommendations.

\section{Datasets and Evaluations}

In this section, we report the datasets and evaluation metrics used by various methods. 
The dataset information comes from the original source or paper.

\subsection{Datasets}

\paragraph{Traditional Recommendation Dataset} 

In Table~\ref{tab:traditional}, we list several traditional recommendation datasets for evaluating model performance.
These datasets provide comprehensive interaction data from various platform, including user-item interactions, timestamps, and review text, enabling the assessment of recommendation models. 
Several state-of-the-art methods have demonstrated their effectiveness using these datasets. 

For instance, the ``Books'' dataset (10.3M users, 4.4M items) from \textbf{Amazon Review data}~\cite{mcauley2015image} has been used to evaluate Agent4Rec~\cite{zhang2024generative} and BiLLP~\cite{shi2024large} performance on large-scale tasks, while the ``Video Games'' dataset (2.8M users, 137.2K items) has validated DRDT~\cite{wang2023drdt} and RAH~\cite{shu2024rah} capabilities. 
The ``Beauty'' dataset (632K users, 112.6K items) has been utilized by IntcRecAgent~\cite{huang2023recommender} and DRDT~\cite{wang2023drdt} to demonstrate their proficiency in recommendation. 
These diverse applications underscore the datasets' crucial role in advancing LLM-powered agent recommender systems and providing a foundation for evaluating various of algorithms.

The \textbf{MovieLens datasets}, introduced by ~\cite{harper2015movielens}, represent another crucial benchmark for evaluating LLM-powered agents recommenders, offering different scales of movie rating data from the MovieLens platform. 
These datasets range from MovieLens-100K (0.9K users, 1.6K items) to MovieLens-20M (138.5K users, 27.3K items), providing researchers with flexibility in testing their methods across different data scales.
Various state-of-the-art approaches have utilized these datasets: FLOW~\cite{cai2024flow} and MACRS~\cite{fang2024multi} have been validated on the smaller MovieLens-100K dataset, while Agent4Rec~\cite{zhang2024generative}, DRDT~\cite{wang2023drdt}, and MACRS~\cite{fang2024multi} have demonstrated their capabilities on MovieLens-1M. 
The larger variants like MovieLens-10M and MovieLens-20M have been employed by InteRecAgent~\cite{huang2023recommender} and RecAgent~\cite{yoon2024evaluating} respectively, showcasing the scalability of their approaches. 
This hierarchical structure of MovieLens datasets makes them particularly valuable for systematically evaluating recommendation algorithms at different scales.

\begin{table*}[t]
\resizebox{0.926\linewidth}{!}{
\begin{tabular}{p{2.5cm}cp{4cm}cccccp{4cm}}
\toprule[1.5pt]
\textbf{Category} & \textbf{Datasets} & \textbf{Reference} & \textbf{Users} & \textbf{Items} & \textbf{Interactions}   & \textbf{Conversations} & \textbf{Turns} & \textbf{Methods} \\
\midrule
\multirow{16}{*}[-5em]{\textbf{\makecell*[c]{Traditional \\ Recommendation \\ Dataset}}}
    & Books & \multirow{8}{*}{\cite{mcauley2015image}} & 10.3M & 
4.4M & 29.5M & - & - & Agent4Rec, BiLLP, RAH, SUBER \\
    & CDs and Vinyl &  & 1.8M &  701.7K  & 4.8M & - & - & AgentCF, KGLA, ToolRec \\
    & Video Games &  & 2.8M & 137.2K & 4.6M  & - & - &  DRDT, RAH, LUSIM \\
    & Beauty &  & 632.0K & 112.6K & 701.5K & - & - & InteRecAgent, DRDT, RecMind \\
    & Clothing &  & 22.6M & 7.2M & 66.0M &- & - & DRDT \\
    & Movies &  &  6.5M & 747.8K & 17.3M &- & - & RAH, LUSIM \\
    & Office Products &  & 7.6M & 710.4K & 12.8M &- & -  & AgentCF \\
    & Music &  & 101.0K & 70.5K & 130.4K &- & - & LUSIM \\
\cmidrule{2-9}

& Movielens-100K & \multirow{4}{*}{\cite{harper2015movielens}} & 0.9K &  
 1.6K & 100K &- & - & FLOW,  MACRS, SUBER \\
    & Movielens-1M &  & 6K &  
 3.7K & 1.0M &- & - & Agent4Rec, RecAgent, DRDT,  MACRS, ToolRec \\
 & Movielens-10M &  & 69.9K &  
 10.6K & 10M &- & - & InteRecAgent \\
  & Movielens-20M &  & 138.5K &  
 27.3K & 20M &- & - &  MACRS, UserSimulator \\
\cmidrule{2-9}

    & Steam & \cite{kang2018self} & 334.7K & 13K & 3.7M &- & - & Agent4Rec, BiLLP, FLOW, InteRecAgent \\
\cmidrule{2-9}
 & Lastfm & \cite{cantador2011second} & 1.2K & 4.6K & 73.5K &- & - & FLOW  \\
\cmidrule{2-9}
 & Yelp & \url{https://www.yelp.com/dataset} & 30.4K & 20.4K & 316.3K &- & - & RecMind, ToolRec, LUSIM \\
\cmidrule{2-9}
 & Anime & \url{https://www.kaggle.com/datasets} & 73.5K & 12.2K & 1.05M &- & - & LUSIM \\
\midrule
\multirow{3}{*}{\textbf{\makecell*[c]{Conversational \\ Recommendation \\ Dataset}}} & ReDial & \cite{li2018towards} & 0.9K & 51.6K & - & 10K & - & UserSimulator, CSHI\\
& Reddit & \cite{he2023large} & 36.2K  & 51.2K & - & 634.4K  & 1.6M & UserSimulator \\
& OpenDialKG & \cite{moon2019opendialkg} & -  & - &  - & 15.6K & 91.2K & CSHI \\
\bottomrule[1.5pt]
\end{tabular}}
\centering
\caption{Summary of Used Experimental Datasets.}
\label{tab:traditional}
\end{table*}

The \textbf{Steam}, \textbf{Lastfm}, \textbf{Anime}, and \textbf{Yelp} datasets provide diverse domain-specific evaluation scenarios for LLM-powered agent recommender systems. 
The Steam dataset, introduced by~\cite{kang2018self}, contains 3.7M interactions between 334.7K users and 13K gaming items, and has been extensively used by methods such as Agent4Rec~\cite{zhang2024generative}, BiLLP~\cite{shi2024large}, FLOW~\cite{cai2024flow}, and InteRecAgent~\cite{huang2023recommender} to validate their effectiveness in game recommendation. 
The Lastfm dataset~\cite{cantador2011second}, focusing on music recommendation, comprises 73.5K interactions from 1.2K users on 4.6K music items, and has been specifically utilized by FLOW~\cite{cai2024flow} to demonstrate its capabilities in the music domain. 
Additionally, the Yelp dataset, containing 316.3K interactions between 30.4K users and 20.4K items, has been employed by RecMind~\cite{wang2024recmind} to evaluate its performance in recommendations.
These domain-specific datasets offer unique evaluation opportunities in specialized recommendation contexts.

% \begin{table*}
% \resizebox{0.7\linewidth}{!}{
% \begin{tabular}{ccccccc}
% \toprule[1.5pt]
%  \textbf{Dataset} & \textbf{Reference} & \textbf{Users} & \textbf{Items} & \textbf{Conversations} & \textbf{Turns}  & \textbf{Methods} \\
% \midrule
% ReDial & \cite{li2018towards} & 0.9K & 51.6K & 10K & - & RecAgent, CSHI\\
% Reddit & \cite{he2023large} & 36.2K  & 51.2K & 634.4K  & 1.6M & RecAgent \\
% OpenDialKG & \cite{moon2019opendialkg} & -  & - & 15.6K & 91.2K & CSHI \\
% \bottomrule[1.5pt]
% \end{tabular}}
% \centering
% \caption{Conversational Recommendation Dataset}
% \label{tab:conversation}
% \end{table*}

\paragraph{Conversational Recommendation Dataset}

In addition to the above traditional recommendation datasets, some works~\cite{cshi}  evaluate the model performance on conversational datasets.
In Table~\ref{tab:traditional}, we list three widely-adopted datasets: \textbf{ReDial}~\cite{li2018towards}, \textbf{Reddit}~\cite{he2023large}, and \textbf{OpenDialKG}~\cite{moon2019opendialkg}. 
The ReDial dataset comprises 11348 multi-turn dialogues involving 6925 movies, where participants engage in seeker-recommender interactions. 
The Reddit dataset is derived from movie recommendation discussions within Reddit communities, where users post recommendation requests and receive responses with movie suggestions, often accompanied by explanatory rationales. 
This extensive dataset encompasses 634392 conversations, 1669720 dialogue turns, 36247 users, and 51203 movies.
CSHI~\cite{cshi} employs ReDial (movie domain, including 10006 dialogues) and OpenDialKG (multiple domains, including 13802 dialogues) for performance evaluation.
UserSimulator~\cite{yoon2024evaluating} evaluates on the Redial and Reddit datasets in a variety of ways, including behavior simulation and memory module believability, etc.
These authentic human-human conversations serve as crucial benchmarks for assessing the model capabilities of LLM-powered agents recommender systems.

It is worth mentioning that considering the agent recommender system based on LLMs, it is necessary to frequently call LLMs or APIs when the model is running. 
In order to save resources and time, some methods sample data from the original dataset for performance evaluation.
For instance, AgentCF~\cite{zhang2024agentcf} randomly samples two subsets (one dense and one sparse), with each subset containing 100 users.
DRDT~\cite{wang2023drdt} randomly samples 200 users from each dataset and uses the target items along with 19 randomly sampled items as the candidate item set.

\subsection{Evaluation}

In Table~\ref{tab:evaluation}, we summary the evaluation metrics used by recent representative methods.

\begin{table*}
\centering
\resizebox{0.9\linewidth}{!}{
\begin{tabular}{p{5cm}p{6cm}p{6cm}}
\toprule[1.5pt]
 \textbf{Category} & \textbf{Metrics} & \textbf{ Methods} \\
\midrule
\multirow{2}{*}[-1em]{Standard Recommendation} & NDCG@K, Recall@K, HR@K, Hit@K, MRR, Acc, F1-Score, MAP & DRDT, RecMind, InteRecAgent, RAH, MACRS, PMS, Agent4Rec, AgentCF, KGLA, FLOW, CSHI, ToolRec, SUBER \\
  & RMSE, MAE, MSE & RecMind \\
\midrule
Language Generation Quality  & BLEU, ROUGE & RecMind, PMS \\
\midrule
Reinforcement Learning  & Rewards & LUSIM, BiLLP, SUBER \\
\midrule
Conversational Efficiency  & Average Turn (AT), Success Rate (SR)  & InteRecAgent, MACRS, CSHI \\
\midrule
\multirow{2}{*}[-1em]{Custom Indicators} & Proactivity, Economy, Explainability, Correctness, Consistency, Efficiency & AutoConcierge \\
& Simulated user behaviors believability, Agent memory believability  & RecAgent \\
\bottomrule[1.5pt]
\end{tabular}}
\centering
\caption{Summary of Used Evaluation Metrics.}
\label{tab:evaluation}
\end{table*}

\paragraph{Standard Recommendation Metrics} Most existing methods employ standard recommendation evaluation metrics to assess model performance. The commonly utilized metrics including Normalized Discounted Cumulative Gain (NDCG@K), Recall@K and Hit Ratio@K (HR@K), etc.
For instance, AgentCF~\cite{zhang2024agentcf} evaluates its performance using NDCG@K and Recall@K on the MovieLens-1M dataset. 
Similarly, DRDT~\cite{wang2023drdt} conducts comprehensive evaluations using Recall@{10,20} and NDCG@{10,20} across multiple datasets including ML-1M, Games, and Luxury datasets.
Hit Ratio@K (HR@K) is another crucial metric for evaluating recommendation performance.
RecMind~\cite{wang2024recmind} employ that for evaluating the recommendation tasks on Amazon Reviews (Beauty) and Yelp datasets.

\paragraph{Language Generation Quality} Some methods~\cite{wang2024recmind} consider the evaluation of language generation quality (e.g., recommendation explanation generation, review summarization), which primarily rely on BLEU and ROUGE metrics.  
BLEU measures the precision of generated text against references, while ROUGE evaluates recall-based similarity, enabling comprehensive assessment of language generation capabilities in recommendation scenarios.
PMS~\cite{pms} utilizes the ROUGE to evaluate the quality of its generated textual recommendations.

\paragraph{Reinforcement Learning Metrics} In evaluating LLM-powered agent recommender systems for long-term engagement, BiLLP~\cite{shi2024large} employs three key metrics adopted from reinforcement learning: trajectory length, average single-round reward, and cumulative trajectory reward. 
Similarly, LUSIM~\cite{lusim} uses the total reward to reflect the overall user engagement during the entire interaction process, and the average reward to represent the average quality of a single recommendation.
These metrics are to evaluate both immediate recommendation quality and long-term engagement effectiveness.

\paragraph{Conversational Efficiency Metrics} Recent research has introduced more comprehensive metrics to evaluate the efficiency of conversational interactions in recommender systems. For instance, MACRS~\cite{fang2024multi} employs key interaction-focused metrics such as Success Rate  (proportion of successful recommendations) and Average Turn (AT) (number of interaction rounds needed to reach a recommendation) per session. 
These metrics assess how effectively the system can understand user preferences and deliver accurate recommendations while minimizing the number of interaction turns.

\paragraph{Custom Indicators} Beyond conventional metrics, some methods~\cite{yoon2024evaluating} propose customized evaluation frameworks.
AutoConcierge~\cite{zeng2024automated} presents six evaluation metrics for task-driven conversational agents: proactivity, economy, explainability, correctness, consistency, and efficiency.
RecAgent~\cite{wang2023user} proposes simulated user behaviors believability and Agent memory believability, to assess the credibility of LLM-simulated user interactions and memory mechanism effectiveness.
These metrics assess system engagement, dialogue efficiency, answer interpretability, response accuracy, requirement fulfillment, and response time, respectively.

In all, these metrics prioritize a holistic understanding of conversational performance, emphasizing balance between efficient recommendation delivery, and maintaining high-quality dialogue throughout the recommendation process.

\section{Related Research Fields}

\paragraph{LLM-powered Recommender Systems} In recent years, recommender systems based on Large Language Models (LLMs) have attracted widespread attention. 
Such systems make full use of the powerful language understanding and generation capabilities of LLMs, bringing a new paradigm to traditional recommender systems.
Most existing methods are primarily designed for rating prediction~\cite{bao2023tallrec} and sequential recommendation~\cite{hou2024large,shao2024ulmrec,zheng2024adapting}.
CoLLM~\cite{zhang2023collm} captures and maps the collaborative information through external traditional models, forming collaborative embeddings used by LLMs. 
LlamaRec~\cite{llamarec} fine-tunes Llama-2-7b for list-wise ranking of the pre-selected items.
However, these methods would face significant limitations: the inability to simulate authentic user behaviors for enhanced personalization, the lack of effective memory mechanisms for long-term context awareness, and the rigid pipeline structure that prevents flexible task decomposition and seamless integration with external tools.

\paragraph{Conversational Recommender Systems}

Conversational recommender systems (CRS) have emerged as a significant research direction in recent years~\cite{jannach2021survey}, which are similar to the LLM-powered agent recommender systems. 
However, traditional methods~\cite{lei2020interactive} have two main drawbacks: attribute-based approaches are limited by rigid dialogue patterns, while generation-based methods suffer from restricted knowledge and poor generalization capabilities of small language models.

\section{Future Directions}

    %\item \textbf{Enhancement of Agent Capabilities}: Current LLM-powered recommendation agents exhibit limitations in behavioral understanding and long-term memory mechanisms. Future research should focus on advancing user intent comprehension, improving knowledge management systems, and enhancing task planning and decision-making capabilities to better accommodate dynamic user preferences.
\paragraph{Optimization of System Architecture} The integration between traditional recommendation methods and LLMs remains insufficient, with challenges in multi-agent collaboration and system interpretability. Future developments should explore flexible architectural designs, enhance agent cooperation efficiency, while ensuring transparency in recommendation.

\paragraph{Refinement of Evaluation Framework} There is a notable absence of unified and comprehensive evaluation standards for accurately measuring dialogue quality and recommendation effectiveness. Future research necessitates the establishment of robust evaluation frameworks, development of novel performance metrics, and consideration of privacy and security concerns in practical applications.

\paragraph{Security Recommender System} \cite{ning2024cheatagent} reveals the vulnerability of LLM-empowered recommender systems to adversarial attacks. In future,  the researchers could develop robust adversarial detection methods, investigate multi-agent defensive architectures, and integrating domain-specific security knowledge into defense mechanisms.
\section{Conclusion}

The integration of LLM-powered agents into recommender systems has emerged as a significant advancement in recent years. 
In this survey, we systematically categorize existing approaches into three paradigms: recommender-oriented, interaction-oriented, and simulation-oriented. 
We comprehensively analyze these methods through a unified four-module architecture and review current datasets and evaluation methodologies.
Finally, we identify three promising directions for future research.

\clearpage

\bibliographystyle{named}
\bibliography{ijcai25}

\begin{thebibliography}{}

\bibitem[\protect\citeauthoryear{Achiam \bgroup \em et al.\egroup }{2023}]{achiam2023gpt}
Josh Achiam, Steven Adler, Sandhini Agarwal, Lama Ahmad, Ilge Akkaya, Florencia~Leoni Aleman, Diogo Almeida, Janko Altenschmidt, Sam Altman, Shyamal Anadkat, et~al.
\newblock Gpt-4 technical report.
\newblock {\em arXiv preprint arXiv:2303.08774}, 2023.

\bibitem[\protect\citeauthoryear{Bao \bgroup \em et al.\egroup }{2023}]{bao2023tallrec}
Keqin Bao, Jizhi Zhang, Yang Zhang, Wenjie Wang, Fuli Feng, and Xiangnan He.
\newblock Tallrec: An effective and efficient tuning framework to align large language model with recommendation.
\newblock In {\em Recsys}, pages 1007--1014, 2023.

\bibitem[\protect\citeauthoryear{Cai \bgroup \em et al.\egroup }{2024}]{cai2024flow}
Shihao Cai, Jizhi Zhang, Keqin Bao, Chongming Gao, and Fuli Feng.
\newblock Flow: A feedback loop framework for simultaneously enhancing recommendation and user agents.
\newblock {\em arXiv preprint arXiv:2410.20027}, 2024.

\bibitem[\protect\citeauthoryear{Cantador \bgroup \em et al.\egroup }{2011}]{cantador2011second}
Iv{\'a}n Cantador, Peter Brusilovsky, and Tsvi Kuflik.
\newblock Second workshop on information heterogeneity and fusion in recommender systems (hetrec2011).
\newblock In {\em Recsys}, pages 387--388, 2011.

\bibitem[\protect\citeauthoryear{Corecco \bgroup \em et al.\egroup }{2024}]{suber}
Nathan Corecco, Giorgio Piatti, Luca~A Lanzend{\"o}rfer, Flint~Xiaofeng Fan, and Roger Wattenhofer.
\newblock An llm-based recommender system environment.
\newblock {\em arXiv preprint arXiv:2406.01631}, 2024.

\bibitem[\protect\citeauthoryear{Fang \bgroup \em et al.\egroup }{2024}]{fang2024multi}
Jiabao Fang, Shen Gao, Pengjie Ren, Xiuying Chen, Suzan Verberne, and Zhaochun Ren.
\newblock A multi-agent conversational recommender system.
\newblock {\em arXiv preprint arXiv:2402.01135}, 2024.

\bibitem[\protect\citeauthoryear{Friedman \bgroup \em et al.\egroup }{2023}]{friedman2023leveraging}
Luke Friedman, Sameer Ahuja, David Allen, Zhenning Tan, Hakim Sidahmed, Changbo Long, Jun Xie, Gabriel Schubiner, Ajay Patel, et~al.
\newblock Leveraging large language models in conversational recommender systems.
\newblock {\em arXiv preprint arXiv:2305.07961}, 2023.

\bibitem[\protect\citeauthoryear{Guo \bgroup \em et al.\egroup }{2024}]{guo2024knowledge}
Taicheng Guo, Chaochun Liu, Hai Wang, Varun Mannam, Fang Wang, Xin Chen, Xiangliang Zhang, and Chandan~K Reddy.
\newblock Knowledge graph enhanced language agents for recommendation.
\newblock {\em arXiv preprint arXiv:2410.19627}, 2024.

\bibitem[\protect\citeauthoryear{Harper and Konstan}{2015}]{harper2015movielens}
F~Maxwell Harper and Joseph~A Konstan.
\newblock The movielens datasets: History and context.
\newblock {\em ACM TIIS}, 5(4):1--19, 2015.

\bibitem[\protect\citeauthoryear{He \bgroup \em et al.\egroup }{2017}]{he2017neural}
Xiangnan He, Lizi Liao, Hanwang Zhang, Liqiang Nie, Xia Hu, and Tat-Seng Chua.
\newblock Neural collaborative filtering.
\newblock In {\em The WebConf}, pages 173--182, 2017.

\bibitem[\protect\citeauthoryear{He \bgroup \em et al.\egroup }{2023}]{he2023large}
Zhankui He, Zhouhang Xie, Rahul Jha, Harald Steck, Dawen Liang, Yesu Feng, Bodhisattwa~Prasad Majumder, Nathan Kallus, and Julian McAuley.
\newblock Large language models as zero-shot conversational recommenders.
\newblock In {\em CIKM}, pages 720--730, 2023.

\bibitem[\protect\citeauthoryear{Hou \bgroup \em et al.\egroup }{2024}]{hou2024large}
Yupeng Hou, Junjie Zhang, Zihan Lin, Hongyu Lu, Ruobing Xie, Julian McAuley, and Wayne~Xin Zhao.
\newblock Large language models are zero-shot rankers for recommender systems.
\newblock In {\em ECIR}, pages 364--381, 2024.

\bibitem[\protect\citeauthoryear{Hu \bgroup \em et al.\egroup }{2008}]{hu2008collaborative}
Yifan Hu, Yehuda Koren, and Chris Volinsky.
\newblock Collaborative filtering for implicit feedback datasets.
\newblock In {\em ICDM}, pages 263--272, 2008.

\bibitem[\protect\citeauthoryear{Huang \bgroup \em et al.\egroup }{2023}]{huang2023recommender}
Xu~Huang, Jianxun Lian, Yuxuan Lei, Jing Yao, Defu Lian, and Xing Xie.
\newblock Recommender ai agent: Integrating large language models for interactive recommendations.
\newblock {\em arXiv preprint arXiv:2308.16505}, 2023.

\bibitem[\protect\citeauthoryear{Jannach \bgroup \em et al.\egroup }{2021}]{jannach2021survey}
Dietmar Jannach, Ahtsham Manzoor, Wanling Cai, and Li~Chen.
\newblock A survey on conversational recommender systems.
\newblock {\em CSUR}, 54(5):1--36, 2021.

\bibitem[\protect\citeauthoryear{Kang and McAuley}{2018}]{kang2018self}
Wang-Cheng Kang and Julian McAuley.
\newblock Self-attentive sequential recommendation.
\newblock In {\em ICDM}, pages 197--206. IEEE, 2018.

\bibitem[\protect\citeauthoryear{Lei \bgroup \em et al.\egroup }{2020}]{lei2020interactive}
Wenqiang Lei, Gangyi Zhang, Xiangnan He, Yisong Miao, Xiang Wang, Liang Chen, and Tat-Seng Chua.
\newblock Interactive path reasoning on graph for conversational recommendation.
\newblock In {\em KDD}, pages 2073--2083, 2020.

\bibitem[\protect\citeauthoryear{Li \bgroup \em et al.\egroup }{2018}]{li2018towards}
Raymond Li, Samira Ebrahimi~Kahou, Hannes Schulz, Vincent Michalski, Laurent Charlin, and Chris Pal.
\newblock Towards deep conversational recommendations.
\newblock {\em NuerIPS}, 31, 2018.

\bibitem[\protect\citeauthoryear{McAuley \bgroup \em et al.\egroup }{2015}]{mcauley2015image}
Julian McAuley, Christopher Targett, Qinfeng Shi, and Anton Van Den~Hengel.
\newblock Image-based recommendations on styles and substitutes.
\newblock In {\em SIGIR}, pages 43--52, 2015.

\bibitem[\protect\citeauthoryear{Moon \bgroup \em et al.\egroup }{2019}]{moon2019opendialkg}
Seungwhan Moon, Pararth Shah, Anuj Kumar, and Rajen Subba.
\newblock Opendialkg: Explainable conversational reasoning with attention-based walks over knowledge graphs.
\newblock In {\em ACL}, pages 845--854, 2019.

\bibitem[\protect\citeauthoryear{Nie \bgroup \em et al.\egroup }{2024}]{hmacrs}
Guangtao Nie, Rong Zhi, Xiaofan Yan, Yufan Du, Xiangyang Zhang, Jianwei Chen, Mi~Zhou, Hongshen Chen, Tianhao Li, Ziguang Cheng, et~al.
\newblock A hybrid multi-agent conversational recommender system with llm and search engine in e-commerce.
\newblock In {\em Recsys}, pages 745--747, 2024.

\bibitem[\protect\citeauthoryear{Ning \bgroup \em et al.\egroup }{2024}]{ning2024cheatagent}
Liang-bo Ning, Shijie Wang, Wenqi Fan, Qing Li, Xin Xu, Hao Chen, and Feiran Huang.
\newblock Cheatagent: Attacking llm-empowered recommender systems via llm agent.
\newblock In {\em KDD}, pages 2284--2295, 2024.

\bibitem[\protect\citeauthoryear{Park \bgroup \em et al.\egroup }{2023}]{park2023generative}
Joon~Sung Park, Joseph O'Brien, Carrie~Jun Cai, Meredith~Ringel Morris, Percy Liang, and Michael~S Bernstein.
\newblock Generative agents: Interactive simulacra of human behavior.
\newblock In {\em AASUIST}, pages 1--22, 2023.

\bibitem[\protect\citeauthoryear{Patil \bgroup \em et al.\egroup }{2023}]{patil2023gorilla}
Shishir~G Patil, Tianjun Zhang, Xin Wang, and Joseph~E Gonzalez.
\newblock Gorilla: Large language model connected with massive apis.
\newblock {\em arXiv preprint arXiv:2305.15334}, 2023.

\bibitem[\protect\citeauthoryear{Schick \bgroup \em et al.\egroup }{2023}]{toolformer}
Timo Schick, Jane Dwivedi-Yu, Roberto Dess{\`\i}, Roberta Raileanu, Maria Lomeli, Eric Hambro, Luke Zettlemoyer, Nicola Cancedda, and Thomas Scialom.
\newblock Toolformer: Language models can teach themselves to use tools.
\newblock In {\em NuerIPS}, volume~36, 2023.

\bibitem[\protect\citeauthoryear{Shao \bgroup \em et al.\egroup }{2024}]{shao2024ulmrec}
Minglai Shao, Hua Huang, Qiyao Peng, and Hongtao Liu.
\newblock Ulmrec: User-centric large language model for sequential recommendation.
\newblock {\em arXiv preprint arXiv:2412.05543}, 2024.

\bibitem[\protect\citeauthoryear{Shen \bgroup \em et al.\egroup }{2024}]{hugginggpt}
Yongliang Shen, Kaitao Song, Xu~Tan, Dongsheng Li, Weiming Lu, and Yueting Zhuang.
\newblock Hugginggpt: Solving ai tasks with chatgpt and its friends in hugging face.
\newblock 36, 2024.

\bibitem[\protect\citeauthoryear{Shi \bgroup \em et al.\egroup }{2024}]{shi2024large}
Wentao Shi, Xiangnan He, Yang Zhang, Chongming Gao, Xinyue Li, Jizhi Zhang, Qifan Wang, and Fuli Feng.
\newblock Large language models are learnable planners for long-term recommendation.
\newblock In {\em SIGIR}, pages 1893--1903, 2024.

\bibitem[\protect\citeauthoryear{Shu \bgroup \em et al.\egroup }{2024}]{shu2024rah}
Yubo Shu, Haonan Zhang, Hansu Gu, Peng Zhang, Tun Lu, Dongsheng Li, and Ning Gu.
\newblock Rah! recsys--assistant--human: A human-centered recommendation framework with llm agents.
\newblock {\em IEEE TCSS}, 2024.

\bibitem[\protect\citeauthoryear{Thakkar and Yadav}{2024a}]{pms}
Param Thakkar and Anushka Yadav.
\newblock Personalized recommendation systems using multimodal, autonomous, multi agent systems.
\newblock {\em arXiv preprint arXiv:2410.19855}, 2024.

\bibitem[\protect\citeauthoryear{Thakkar and Yadav}{2024b}]{thakkar2024personalized}
Param Thakkar and Anushka Yadav.
\newblock Personalized recommendation systems using multimodal, autonomous, multi agent systems.
\newblock {\em arXiv preprint arXiv:2410.19855}, 2024.

\bibitem[\protect\citeauthoryear{Wang \bgroup \em et al.\egroup }{2023a}]{wang2023user}
Lei Wang, Jingsen Zhang, Hao Yang, Zhiyuan Chen, Jiakai Tang, Zeyu Zhang, Xu~Chen, Yankai Lin, Ruihua Song, Wayne~Xin Zhao, et~al.
\newblock User behavior simulation with large language model based agents.
\newblock {\em arXiv preprint arXiv:2306.02552}, 2023.

\bibitem[\protect\citeauthoryear{Wang \bgroup \em et al.\egroup }{2023b}]{wang2023drdt}
Yu~Wang, Zhiwei Liu, Jianguo Zhang, Weiran Yao, Shelby Heinecke, and Philip~S Yu.
\newblock Drdt: Dynamic reflection with divergent thinking for llm-based sequential recommendation.
\newblock {\em arXiv preprint arXiv:2312.11336}, 2023.

\bibitem[\protect\citeauthoryear{Wang \bgroup \em et al.\egroup }{2024a}]{wang2024survey}
Lei Wang, Chen Ma, Xueyang Feng, Zeyu Zhang, Hao Yang, Jingsen Zhang, Zhiyuan Chen, Jiakai Tang, Xu~Chen, Yankai Lin, et~al.
\newblock A survey on large language model based autonomous agents.
\newblock {\em Frontiers of Computer Science}, 18(6):186345, 2024.

\bibitem[\protect\citeauthoryear{Wang \bgroup \em et al.\egroup }{2024b}]{wang2024recmind}
Yancheng Wang, Ziyan Jiang, Zheng Chen, Fan Yang, Yingxue Zhou, Eunah Cho, Xing Fan, Yanbin Lu, Xiaojiang Huang, and Yingzhen Yang.
\newblock Recmind: Large language model powered agent for recommendation.
\newblock In {\em Findings of NAACL}, pages 4351--4364, 2024.

\bibitem[\protect\citeauthoryear{Wang \bgroup \em et al.\egroup }{2024c}]{wang2024macrec}
Zhefan Wang, Yuanqing Yu, Wendi Zheng, Weizhi Ma, and Min Zhang.
\newblock Macrec: A multi-agent collaboration framework for recommendation.
\newblock In {\em SIGIR}, pages 2760--2764, 2024.

\bibitem[\protect\citeauthoryear{Wu \bgroup \em et al.\egroup }{2022}]{wu2022survey}
Le~Wu, Xiangnan He, Xiang Wang, Kun Zhang, and Meng Wang.
\newblock A survey on accuracy-oriented neural recommendation: From collaborative filtering to information-rich recommendation.
\newblock {\em IEEE TKDE}, 35(5):4425--4445, 2022.

\bibitem[\protect\citeauthoryear{Yao \bgroup \em et al.\egroup }{2023}]{react}
Shunyu Yao, Jeffrey Zhao, Dian Yu, Nan Du, Izhak Shafran, Karthik~R Narasimhan, and Yuan Cao.
\newblock React: Synergizing reasoning and acting in language models.
\newblock In {\em ICLR}, 2023.

\bibitem[\protect\citeauthoryear{Yoon \bgroup \em et al.\egroup }{2024}]{yoon2024evaluating}
Se-eun Yoon, Zhankui He, Jessica~Maria Echterhoff, and Julian McAuley.
\newblock Evaluating large language models as generative user simulators for conversational recommendation.
\newblock {\em arXiv preprint arXiv:2403.09738}, 2024.

\bibitem[\protect\citeauthoryear{Yue \bgroup \em et al.\egroup }{2023}]{llamarec}
Zhenrui Yue, Sara Rabhi, Gabriel de Souza~Pereira Moreira, Dong Wang, and Even Oldridge.
\newblock Llamarec: Two-stage recommendation using large language models for ranking.
\newblock {\em arXiv preprint arXiv:2311.02089}, 2023.

\bibitem[\protect\citeauthoryear{Zeng \bgroup \em et al.\egroup }{2024}]{zeng2024automated}
Yankai Zeng, Abhiramon Rajasekharan, Parth Padalkar, Kinjal Basu, Joaqu{\'\i}n Arias, and Gopal Gupta.
\newblock Automated interactive domain-specific conversational agents that understand human dialogs.
\newblock In {\em ISPADL}, pages 204--222, 2024.

\bibitem[\protect\citeauthoryear{Zhang \bgroup \em et al.\egroup }{2023}]{zhang2023collm}
Yang Zhang, Fuli Feng, Jizhi Zhang, Keqin Bao, Qifan Wang, and Xiangnan He.
\newblock Collm: Integrating collaborative embeddings into large language models for recommendation.
\newblock {\em arXiv preprint arXiv:2310.19488}, 2023.

\bibitem[\protect\citeauthoryear{Zhang \bgroup \em et al.\egroup }{2024a}]{zhang2024generative}
An~Zhang, Yuxin Chen, Leheng Sheng, Xiang Wang, and Tat-Seng Chua.
\newblock On generative agents in recommendation.
\newblock In {\em SIGIR}, pages 1807--1817, 2024.

\bibitem[\protect\citeauthoryear{Zhang \bgroup \em et al.\egroup }{2024b}]{Rec4Agentverse}
Jizhi Zhang, Keqin Bao, Wenjie Wang, Yang Zhang, Wentao Shi, Wanhong Xu, Fuli Feng, and Tat-Seng Chua.
\newblock Prospect personalized recommendation on large language model-based agent platform.
\newblock {\em arXiv preprint arXiv:2402.18240}, 2024.

\bibitem[\protect\citeauthoryear{Zhang \bgroup \em et al.\egroup }{2024c}]{zhang2024agentcf}
Junjie Zhang, Yupeng Hou, Ruobing Xie, Wenqi Sun, Julian McAuley, Wayne~Xin Zhao, Leyu Lin, and Ji-Rong Wen.
\newblock Agentcf: Collaborative learning with autonomous language agents for recommender systems.
\newblock In {\em The WebConf}, pages 3679--3689, 2024.

\bibitem[\protect\citeauthoryear{Zhang \bgroup \em et al.\egroup }{2024d}]{lusim}
Zijian Zhang, Shuchang Liu, Ziru Liu, Rui Zhong, Qingpeng Cai, Xiangyu Zhao, Chunxu Zhang, Qidong Liu, and Peng Jiang.
\newblock Llm-powered user simulator for recommender system.
\newblock {\em arXiv preprint arXiv:2412.16984}, 2024.

\bibitem[\protect\citeauthoryear{Zhao \bgroup \em et al.\egroup }{2024}]{toolrec}
Yuyue Zhao, Jiancan Wu, Xiang Wang, Wei Tang, Dingxian Wang, and Maarten De~Rijke.
\newblock Let me do it for you: Towards llm empowered recommendation via tool learning.
\newblock In {\em SIGIR}, pages 1796--1806, 2024.

\bibitem[\protect\citeauthoryear{Zheng \bgroup \em et al.\egroup }{2024}]{zheng2024adapting}
Bowen Zheng, Yupeng Hou, Hongyu Lu, Yu~Chen, Wayne~Xin Zhao, Ming Chen, and Ji-Rong Wen.
\newblock Adapting large language models by integrating collaborative semantics for recommendation.
\newblock In {\em ICDE}, pages 1435--1448. IEEE, 2024.

\bibitem[\protect\citeauthoryear{Zhu \bgroup \em et al.\egroup }{2024a}]{cshi}
Lixi Zhu, Xiaowen Huang, and Jitao Sang.
\newblock A llm-based controllable, scalable, human-involved user simulator framework for conversational recommender systems.
\newblock {\em arXiv preprint arXiv:2405.08035}, 2024.

\bibitem[\protect\citeauthoryear{Zhu \bgroup \em et al.\egroup }{2024b}]{zhu2024recommender}
Xi~Zhu, Yu~Wang, Hang Gao, Wujiang Xu, Chen Wang, Zhiwei Liu, Kun Wang, Mingyu Jin, Linsey Pang, Qingsong Wen, et~al.
\newblock Recommender systems meet large language model agents: A survey.
\newblock {\em SSRN 5062105}, 2024.

\end{thebibliography}

\clearpage

\end{document}